\documentclass[RNAAS,twocolumn]{aastex701}
\usepackage{graphicx}

\begin{document}

\title{ColdPress: Efficient Quantile-Based Compression of Photometric Redshift PDFs}

\author[0000-0002-4237-5500]{Antonio Hern\'an-Caballero}
\affiliation{Centro de Estudios de F\'isica del Cosmos de Arag\'on (CEFCA), Plaza San Juan, 1, E-44001 Teruel, Spain}
\email[show]{ahernan@cefca.es}

\begin{abstract}
\texttt{ColdPress} is a Python module that compresses photometric redshift probability distribution functions (PDFs) by encoding quantiles of their cumulative distribution. For a fixed packet size (the default is 80 bytes per PDF), \texttt{ColdPress} attains a reconstruction accuracy comparable to the sparse-basis representation method implemented in the \texttt{pdf\_storage} module of \citet{CarrascoKind2014}, yet reduces the computational cost by a factor of $\sim$7000. I describe the implementation and quantify its compression speed and reconstruction accuracy in comparison to \texttt{pdf\_storage} for real-life PDFs from two different photometric redshift codes. 
\texttt{ColdPress} is free software, available at \url{https://github.com/ahc-photoz/coldpress-project}.
\end{abstract}

\section{Introduction}

Photometric redshift PDFs provide the full uncertainty distribution for the redshift of each astronomical source. This is critical for estimating errors in redshift-dependent quantities and for ensemble analyses like \(N(z)\) estimation and clustering. However, storing and distributing billions of PDFs from large-area surveys (e.g. Euclid, LSST, J-PAS) presents significant challenges. 

\citet{CarrascoKind2014} introduced \texttt{pdf\_storage}, a PDF compression method using sparse-basis representation with Orthogonal Matching Pursuit over a large dictionary of Gaussian and Voigt profiles. This method provides high compression ratios and accurate PDF reconstruction. Unfortunately, optimizing the sparse-basis representation coefficients has a high computational cost per PDF. Additionally, \texttt{pdf\_storage} requires P($z$) to be sampled on a predefined redshift grid. While binned PDFs is the natural output of template-fitting photo-$z$ codes, regression methods (including machine learning algorithms) often deterministically output a single redshift for any given input photometry, with PDFs generated via Monte Carlo sampling of the fluxes according to their uncertainties \citep{Tanaka2018}. For such PDFs, constructing binned representations on a fixed grid requires many samples to overcome shot noise. 

A more efficient approach is to store the redshifts \{$z_i$\} of the quantiles \{$q_i$\} of the cumulative distribution function (CDF; see \citealp{Malz2018} for a discussion). Quantiles naturally sample in finer detail redshift intervals with higher probability density. Furthermore, no redshift grid is needed, allowing sources with widely different redshift ranges (like bright galaxies and QSOs) to be stored in the same table.

\texttt{ColdPress} is a Python module that efficiently compresses PDFs by encoding the redshift separation between consecutive quantiles of the CDF. \texttt{ColdPress} encoding reduces size by $\sim$3.75 times compared to storing quantiles as 32-bit floating-point numbers, with no significant accuracy loss. For binned PDFs, \texttt{ColdPress} achieves comparable reconstruction accuracy to \texttt{pdf\_storage} at the same compression ratio while decreasing the computational cost by a factor $\sim$7000.

\section{Implementation details}

Efficient storage requires all PDFs to be encoded as packets of uniform byte size. The default is 80 bytes: 5 for the header and 75 for the payload. 
The first byte in the header indicates the redshift precision of the encoding, $\epsilon$, in units of 10$^{-5}$. This limits the accuracy of the recovered $z_i$ to between 1$\times$10$^{-5}$ ($\sim$3 km/s) and 255$\times$10$^{-5}$ ($\sim$750 km/s). 
The next four bytes encode the first ($z_0$) and last ($z_{n-1}$) quantile redshifts as 16-bit unsigned integers, calculated as the nearest integer to 5000($z$+0.01). This method provides a 0.0002 precision for redshifts up to $z$=13, while the 0.01 offset allows negative velocities (to -3000 km/s) to be stored as positive integers.

The payload encodes the differences $\Delta_i$ = ($z_i$ - $z_{i-1}$)/$\epsilon$ rounded to the nearest integer. If $\Delta_i$ $\le$ 254, a single byte is used. For larger values, a special 3-byte encoding is used in which the first byte is 255 to flag the big jump, and the next two bytes encode $\Delta_i$. A large enough $\epsilon$ to fit most of the $\Delta_i$ in a single byte is important, since big jumps are expensive. For a packet of length $L$ and $n$ quantiles, the maximum number of big jumps is $m$ = ($L$ - $n$ - 3)/2. 
In practice, \texttt{ColdPress} determines the optimal combination of quantiles ($n$) and large jumps ($m$) for each PDF. For an 80-byte packet, most real-life PDFs are encoded with zero to three large jumps, resulting in 71 to 77 quantiles. 

In addition to the core functions for encoding quantiles into byte packets and decoding them back, \texttt{Coldpress} includes functions for converting binned PDFs into quantile CDFs and vice versa, summary statistics from the quantiles (mode, mean, confidence intervals, odds, etc.), and more. Further information is available in the documentation at the \texttt{ColdPress} repository\footnote{\url{https://github.com/ahc-photoz/coldpress-project}}.

\section{Comparison with Sparse-Basis Compression}

I compare the efficiency of \texttt{ColdPress} and \texttt{pdf\_storage} using PDFs from the Hyper-SuprimeCam Subaru Strategic Survey (HSC-SSP) Public Data Release 3 \citep[PDR3;][]{Aihara2022}. The PDFs are generated with two different codes: (a) \texttt{Mizuki} \citep{Tanaka2015,Tanaka2018}, a SED-fitting code that evaluates P(z) in a grid from $z$=0 to $z$=7.0 in steps of 0.01, and (b) \texttt{DEmP} \citep{Hsieh2014,Tanaka2018}, which uses an empirical, nearest-neighbor + polynomial fitting approach combined with Monte Carlo sampling of P(z). \texttt{DEmP} samples are binned from $z$=0 to $z$=6.0 in steps of 0.01. 

I selected all sources in tract 16821 of HSC-SSP PDR3, which partially covers the AEGIS field. The \texttt{DEmP} catalog contains 33,491 sources, while the \texttt{Mizuki} catalog contains 17,439 (all in common with \texttt{DEmP}).
 
Each PDF from \texttt{DEmP} (\texttt{Mizuki}) takes 2404 (2804) bytes uncompressed. Encoding in 80 bytes implies a compression ratio of 30 (35).
Running on a modern laptop, the compression of 33,491 \texttt{DEmP} PDFs took 11.5 CPU hours with \texttt{pdf\_storage} but just 5.1 CPU seconds with \texttt{ColdPress}, a factor of $\sim$8100 increase in speed. For \texttt{Mizuki} PDFs, the speed gain was a factor of $\sim$7300.

I define the error in the reconstruction of the binned PDF from its compressed representation as
\begingroup
  \setlength{\abovedisplayskip}{3pt}
  \setlength{\belowdisplayskip}{3pt}
  \setlength{\abovedisplayshortskip}{3pt}
  \setlength{\belowdisplayshortskip}{3pt}
  \[
    \zeta = \sum_{i=1}^{N}
      \bigl|F_{\rm orig}(z_i) - F_{\rm rec}(z_i)\bigr|
  \]
\endgroup

\noindent where $F_{\mathrm{orig}}(z_i)$ is the original CDF at bin $z_i$, $F_{\mathrm{rec}}(z_i)$ is the reconstructed CDF at $z_i$, and $N$ is the number of redshift bins in the PDF. 

\begin{figure*}
\centering
\includegraphics[width=17cm]{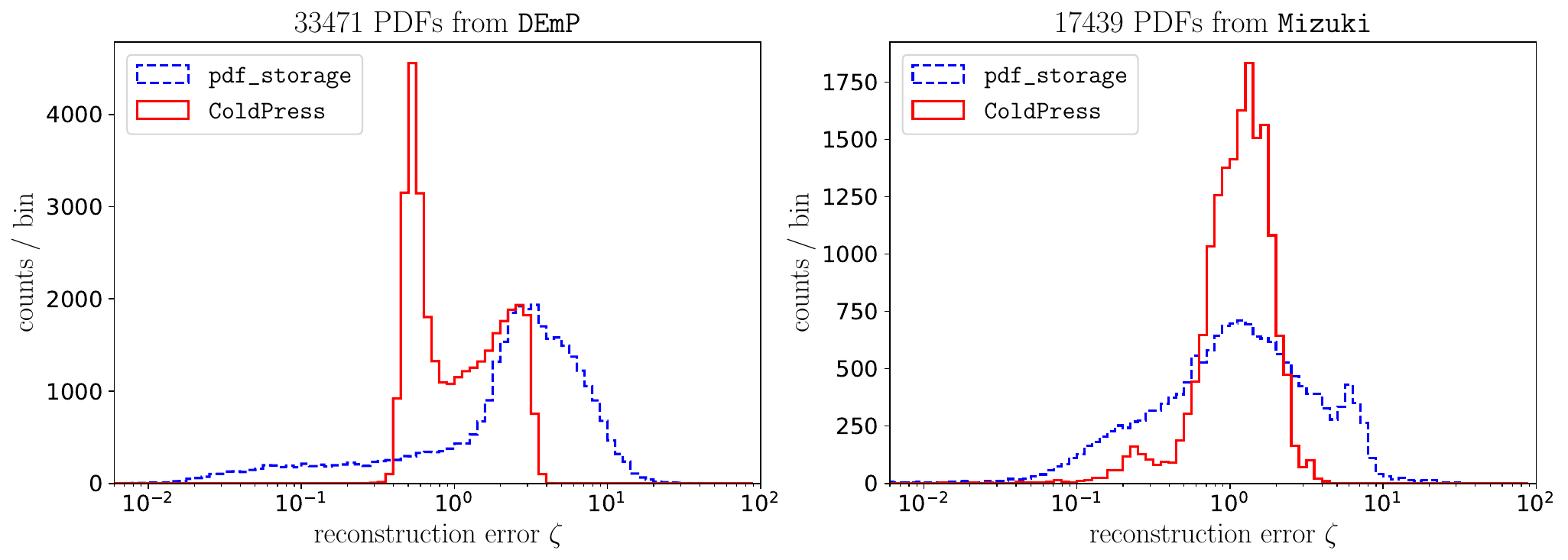}
\includegraphics[width=17cm]{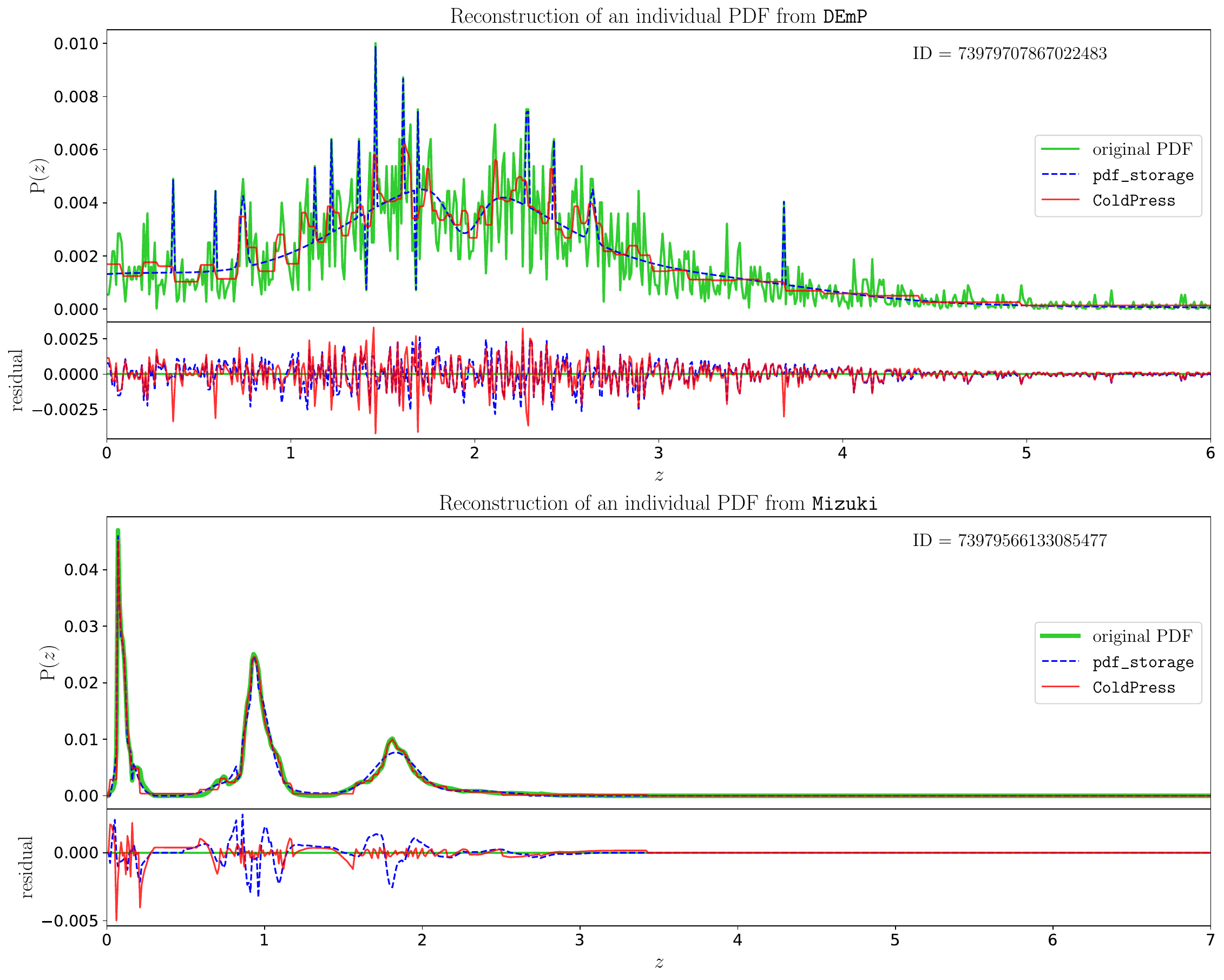}
\caption{Comparison of the accuracy of PDF reconstructions with \texttt{pdf\_storage} and \texttt{ColdPress}. The top panels show the distribution of the reconstruction error for individual sources on PDFs generated via Monte Carlo sampling (\texttt{DEmP} code; left) or SED-fitting with grid sampling (\texttt{Mizuki} code; right). The middle and bottom panels show the original PDF, its reconstructions with \texttt{pdf\_storage} and \texttt{ColdPress}, and their residuals for two representative PDFs.}
\label{fig:comparison}
\end{figure*}

Figure \ref{fig:comparison} compares the PDF reconstruction of \texttt{pdf\_storage} and \texttt{ColdPress}. The top panels show that while both methods have comparable median values of $\zeta$, \texttt{pdf\_storage} has a wider performance range, whereas \texttt{ColdPress} is more consistent.
Examples of PDF reconstructions are shown in the middle and bottom panels of Fig. \ref{fig:comparison}. Both \texttt{pdf\_storage} and \texttt{ColdPress} struggle to reproduce the noisy PDFs from \texttt{DEmP}. \texttt{pdf\_storage} emphasizes the strongest spikes and models the rest as a very smooth continuum, while \texttt{ColdPress} smooths the entire PDF to a lower extent. For \texttt{Mizuki} PDFs, \texttt{ColdPress} reproduces the peaks better, while \texttt{pdf\_storage} is more accurate in the wings.

\begin{thebibliography}{}
\bibitem[Aihara et al.(2022)]{Aihara2022} Aihara, H., AlSayyad, Y., Ando, M., et al.\ 2022, \emph{PASJ}, 74, 247
\bibitem[Carrasco-Kind \& Brunner(2014)]{CarrascoKind2014} Carrasco-Kind, M., \& Brunner, R.~J.\ 2014, \emph{MNRAS}, 441, 3550
\bibitem[Hsieh \& Yee(2014)]{Hsieh2014}Hsieh, B. C., \& Yee, H. K. C. 2014, \emph{ApJ}, 792, 102
\bibitem[Malz et al.(2018)]{Malz2018} Malz, A.~I., Marshall, P.~J., DeRose, J., et al.\ 2018, \emph{AJ}, 156, 35
\bibitem[Tanaka(2015)]{Tanaka2015} Tanaka, M.\ 2015, \emph{ApJ}, 801, 20
\bibitem[Tanaka et al.(2018)]{Tanaka2018} Tanaka, M., Coupon, J., Hsieh, B.-C., et al.\ 2018, \emph{PASJ}, 70, S9
\end{thebibliography}
\end{document}